# GPTutor: an open-source AI pair programming tool alternative to Copilot

EASON CHEN, Carnegie Mellon University and Bucket Protocol

RAY HUANG, JUSTA LIANG, DAMIEN CHEN, and PIERCE HUNG, Bucket Protocol

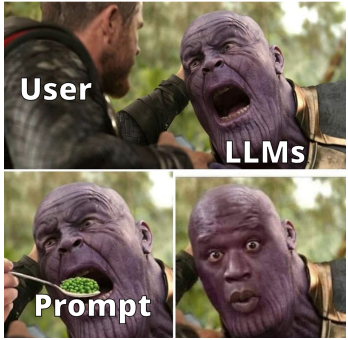 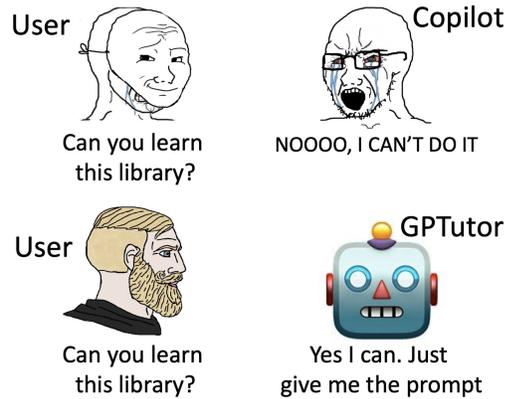

Fig. 1. Memes to illustrate the goal of GPTutor: enable users to customize prompts to improve the output generated by LLMs.

This paper presents the latest progress of GPTutor: a ChatGPT-powered programming tool extension in Visual Studio Code. The emergence of Large Language Models (LLMs) has improved software development efficiency, but their performance can be hindered by training data limitations and prompt design issues. Existing LLM development tools often operate as black boxes, with users unable to view the prompts used and unable to improve performance by correcting prompts when errors occur. To address the aforementioned issues, GPTutor was introduced as an open-source AI pair programming tool, offering an alternative to Copilot. GPTutor empowers users to customize prompts for various programming languages and scenarios, with support for 120+ human languages and 50+ programming languages. Users can fine-tune prompts to correct the errors from LLM for precision and efficient code generation. At the end of the paper, we underscore GPTutor's potential through examples, including demonstrating its proficiency in interpreting and generating Sui-Move, a newly introduced smart contract language, using prompt engineering.

## 1 INTRODUCTION

Thanks to Large Language Models [1], the efficiency of modern software engineers in developing programs has significantly improved. Software engineers frequently employ various AI tools to assist development, such as Copilot, CodeX and ChatGPT. These tools are primarily used for two key purposes: code explanation and generation[7]. "Code Explanation" pertains to providing clear and understandable explanations for given sections of code. "Code generation" refers to AI generating the corresponding code based on developer instructions or code. For example, AI may generates the remaining incomplete portions of code based on the code already written by the developer.

These LLM AI Pair Programming tools are generally effective in handling most scenarios [9]. Nevertheless, it's important to note that large language models such as ChatGPT are only equipped with knowledge up to 2021. As a result, developers might face challenges when dealing with recently introduced programming languages or libraries. For example, introduced in 2022, Sui-Network uses Sui-Move as the programming language for its smart contract logic. As illustrated in Figure 2, Pair programming tools powered by LLM AI struggle to provide significant assistance in the development of projects involving unfamiliar libraries like Sui-Move, due to their training data limitations.

Fortunately, prompt Engineering can easily address these issues (Figure 1). For instance, users can provide background in the prompt, such as mentioning that "Move language has syntax similar to Rust", before asking ChatGPT to explain Sui-Move code. With this context, ChatGPT can effectively explain Sui-Move code. Furthermore, by providing Sui-Move examples in the prompt accordingly, ChatGPT can successfully generate or modify Sui-Move smart contracts.

Unfortunately, development tools like Copilot are black boxes, lacking the flexibility to customize prompts or correct errors as per one's requirements [9] (Figure 1). Moreover, pasting code into ChatGPT and selecting the appropriate prompt can be quite cumbersome. This is why we are building GPTutor, an open-source AI pair programming tool that aims to be an alternative to Copilot.

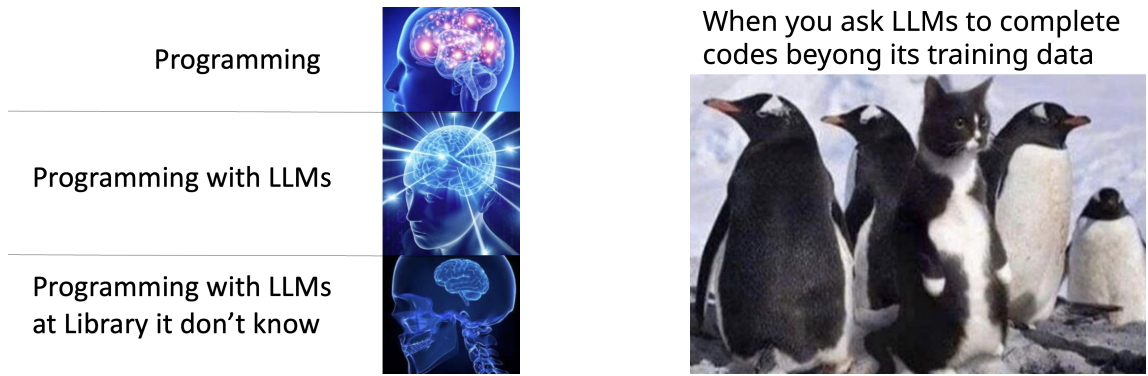

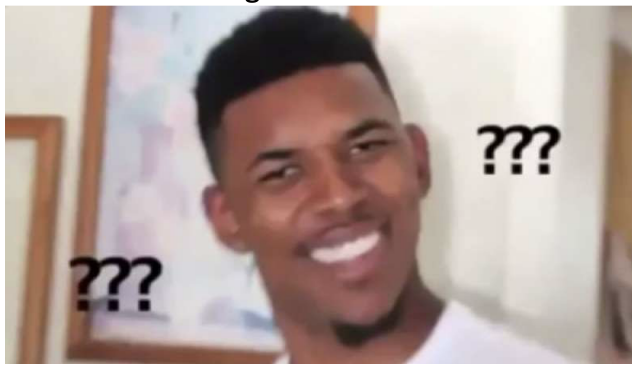

Fig. 2. Memes to illustrate the issue.

```
public entry fun mint(
    treasury_cap: &mut TreasuryCap<MANAGED>,
    amount: u64,
    recipient: address,
    ctx: &mut TxContext
) {
    coin::mint_and_transfer( c: treasury_cap,  amount: 100, recipient, ctx)
}
```

(Explain)

The selected Move code is a function called `mint` that is part of the `managed` module in a smart contract. This function is declared as a public entry point, meaning it can be called externally by other modules or accounts.

The `mint` function takes four parameters:

- `treasury_cap`: A mutable reference to a `TreasuryCap` object. `TreasuryCap` represents the total supply of the managed coin and is specific to the `MANAGED` coin type.
- `amount`: An unsigned 64-bit integer representing the number of coins to be minted.
- `recipient`: An address representing the account that will receive the minted coins.
- `ctx`: A mutable reference to the transaction context, which provides about the current transaction.

Within the `mint` function, there is a call to another function called `coin::mint_and_transfer`. This function is responsible for both minting new coins and transferring them to the specified recipient.

(Comment)

```
/// This function allows the trusted manager
to mint new coins and transfer them to a
recipient.
/// It takes the mutable reference of the
`treasury_cap` representing the treasury cap
of the managed coin
/// the `amount` of coins to be minted,
/// the `recipient` address where the minted
coins will be transferred,
/// and the mutable reference of the
transaction context `ctx`.
```

(Review)

**Security Concerns:**

- In the `mint` function, the `amount` parameter is not being used in the `mint_and_transfer` function call. Instead, a fixed value of `100` is being passed as the amount. This seems like a bug or oversight. To address this, I have modified the code to use the `amount` argument passed to the `mint` function.

Fig. 3. Examples for using GPTutor for Explain, Comment, and Code Review on a Sui-Move code snips

## 2 GPTUTOR SERVICE DESIGN

Currently, GPTutor is available at the Visual Studio Code Extension marketplace with over a thousand downloads [1]. Users can use it with their own OpenAI API Key. GPTutor supports input and output in over 120 human languages and supports more than fifty programming languages. Users can customize GPTutor's prompts for specific languages to obtain more precise explanations or generations. For example, as shown in Figure 3, developers can set up prompts for GPTutor to explain, comments, and even perform code review for their code. Moreover, as shown in Figure 5 users can specify what CSS library and themes they want to use and then ask the GPTutor to rewrite the HTML classes to fit the instructions. For more about the design of GPTutor, please check [4].

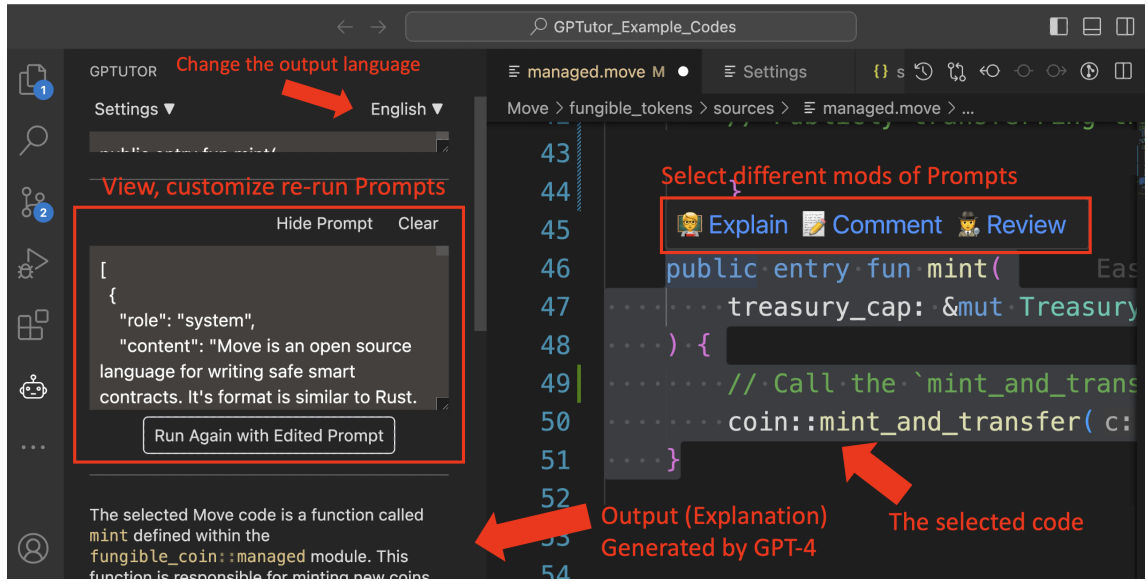

Fig. 4. GPTutor's user interface and it's annotation

## 3 CUSTOMIZE PROMPTS

With GPTutor, users can customize their prompts for various programming languages and development scenarios and easily switch between different prompts as needed (Figure 4). Leveraging the capabilities of GPT-3.5 and GPT-4, GPTutor offers support for customized Prompts in LangChain Template format. Notably, GPTutor not only supports inputting code from the active window as the prompt, but it can also pick the source code behind the function chosen by the user as a prompt. By these kinds of in-depth analyses, GPTutor can generate even more precise outputs than using vanilla ChatGPT and Copilot[4].

Please check this demo video for functions of GPTutor. While utilizing GPTutor, the prompts it employs are completely transparent. Users can easily view the current prompt with a single click and instantly edit to see how different prompts can produce varying output. If users are satisfied with the changes, they can save the modified prompt in the configuration as personalized prompts. Finally, users are encouraged to submit Pull Requests to share the prompts they've used to improve the overall GPTutor Community.

We have specifically tailored GPTutor's prompts to enhance its ability to explain and generate Sui-Move. This customization is designed to aid developers in swiftly comprehending Sui-Move development and serves as an example of how to tailor prompts for a particular programming language. For example, as shown in Figure 3, developers can set up prompts to use GPTutor on Sui Sui-Move, a language beyond ChatGPT's training data, to explain Sui-Move, generate comments for Sui-Move code, and even perform code review for their Sui-Move smart contract. Moreover, by including the Sui-Move Fungible Coin Smart Contract Template in the prompt as a reference, GPTutor can accurately generate and modify Sui-Move smart contracts code related to Fungible Coins. This is aimed at helping developers understand the workings of Sui-Move contracts and expedite the development of their first Fungible Coin smart contract.

---

[1]GPTutor Download Link: https://marketplace.visualstudio.com/items?itemName=gptutor.gptutor

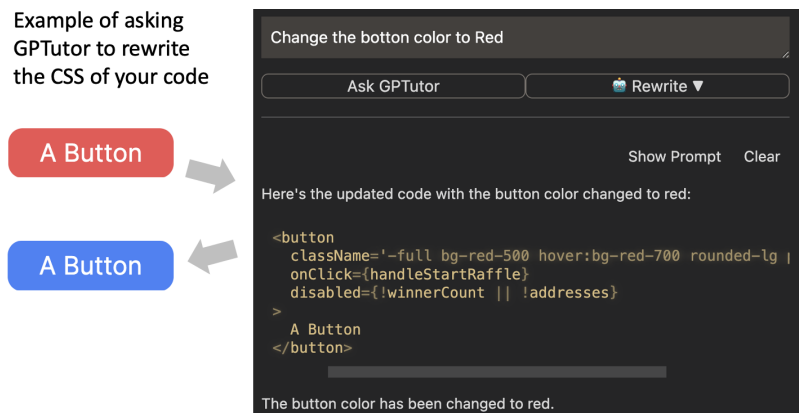

Fig. 5. Example of using GPTutor to change to button CSS styles with instructions.

## 4 FUTURE WORKS

In the future, we hope to further enhance GPTutor's capabilities with prompt engineering, enabling it to support a wider range of contexts and even allowing it to switch prompts autonomously based on the context. Additionally, we aim to improve the user experience when it comes to editing prompts in GPTutor, such as generating multiple responses [2] to help users compare which prompt works best [6]. Moreover, we're also interested in employing NLP Factors to identify the most appropriate explanations and recommendations [3]. Finally, we also intend to conduct practical research on GPTutor's Tutur capabilities, exploring how GPTutor can assist developers in efficiently adopting emerging technologies, such as assisting in the conversion of natural language documents to domain-specific language [5, 8].

## 5 CALL FOR CONTRIBUTION

GPTutor welcomes contributions for custom prompts, especially those related to new programming languages or libraries. You're more than welcome to submit Issues and Pull Requests on GitHub https://github.com/GPTutor/gptutor-extension.

## 6 CONCLUSION

In this paper, we discuss the limitations of using AI Paired Programming with LLMs for development and explanation. We then introduce GPTutor, a tool that allows users to customize prompts to address this issue. We outline the current progress of GPTutor and future development plans.

## ACKNOWLEDGMENTS

This work was supported by the Sui Foundation. This work is presented at the Workshop of Supporting User Engagement in Testing, Auditing, and Contesting AI at The 26th ACM Conference On Computer-Supported Cooperative Work And Social Computing. Funding to attend the conference was provided by the CMU GSA/Provost Conference Funding.